\documentclass[aps,prl,twocolumn]{revtex4-1}

\usepackage{graphicx}% Include figure files
\usepackage{dcolumn}% Align table columns on decimal point
\usepackage{bm}% bold math
\usepackage{ textcomp }
\usepackage{color}
\usepackage{amsmath}
\usepackage{amssymb}
\usepackage{xcolor}
\usepackage{hyperref}
\usepackage{easyReview}
\usepackage{mathdots}
\usepackage{float}
\usepackage{lineno}
\usepackage{array}
\usepackage{xcolor,colortbl}
\definecolor{LightBlue}{rgb}{0.8,0.8,0.8}

\begin{document}
\title{Reinvestigation of the homogeneous spin model in YbMgGaO$_4$}
\date{\today}

\author{Shaozhi Li}
%\email{szli@umich.edu}
\affiliation{Department of Physics, University of Michigan, Ann Arbor, Michigan 48109, USA}
\affiliation{Materials Science and Technology Division, Oak Ridge National Laboratory, Oak Ridge, Tennessee 37831, USA}

\begin{abstract}
Motivated by a recent inelastic neutron scattering experiment on $\mathrm{YbMgGaO}_4$~\cite{William2019}, I reinvestigate the homogeneous spin model on the triangular lattice. Using the cluster mean-field theory and exact diagonalization, I studied the phase diagram and magnetic-field-induced phase transitions. The results show that the magnetic field can induce the phase transition from the spin liquid state (or the stripe state) to the $120^{\circ}$ antiferromagnetic state.
These phase transitions are suppressed by the next-nearest neighbor exchange interaction $J_2/J_1$ and vanish as $J_2/J_1>0.11$. I analyze a parameter space at $J_2/J_1=0.1$, where a field-induced phase transition can occur, and find that the deviation of the theoretical spin excitation energy from the experimental data is only about $0.054$. These results imply that an effective homogeneous spin model still applies to $\mathrm{YbMgGaO}_4$.
\end{abstract}
\maketitle
%{\it Introduction ---} 
\section{Introduction}
Quantum spin liquids refer to a novel state where spins do not form an ordered pattern down to zero temperature in spite of strong spin-spin interactions~\cite{Broholmeaay0668}. This state has been found in a variety of materials and models with frustrated geometries, including triangular and 
kagome lattices.~\cite{Balents2010nature,ZhouPRL2012,RawlPRB2017,Gardner2010RMP,MENDELS2016455,Zhou2017RMP}. To date, the quantum spin liquid is being actively explored due to the great potential for applications in quantum computing, telecommunications, and spintronics~\cite{Savary_2016}.

Recently, $\mathrm{YbMgGaO}_4$ has been attracting much attention as a candidate for the spin liquid state~\cite{Li2015SCI,Yuesheng2015PRL,Yuesheng2016PRL,Shen2016nature,Xu2016PRL,shen2018naturecom,LiPRB2018,Paddison2017,Yuesheng2017PRL,Majumder2020PRR,Yuesheng2017NC}.
In $\mathrm{YbMgGaO}_4$, the magnetic ions, $\mathrm{Yb}^{3+}$, form a perfect triangular lattice. The spin-orbit coupling (SOC) and the crystal field lead to a Kramers' doublet for the $\mathrm{Yb}^{3+}$ ion, which is described by
an effective spin-1/2 moment~\cite{LiPRB2018}. 
The interaction between the effective spin-1/2 moments is anisotropic due to the spin-orbit-entangled nature of the Kramers' doublets~\cite{Maksimov2019PRX}.
Besides, the mixing of Mg and Ga in $\mathrm{YbMgGaO}_4$ leads to disorder~\cite{Yuesheng2016PRL}. Although this disordering effect has been discussed in several studies~\cite{WuPRB2019,ZhuPRL2017,Parker2018PRB,Kimchi2018PRX}, the mainstream view is that an effective homogeneous model can capture the spin properties in $\mathrm{YbMgGaO}_4$~\cite{dongPRB2016,LiPRB2018,Luo2017PRB,Parker2018PRB} because the extensively used A~\cite{Paddison2017}, B~\cite{LiPRB2018}, and C~\cite{ZhangPRX} models can reproduce the inelastic neutron scattering (INS) data at high fields~\cite{ZhangPRX,Supplement}.

This idea has been challenged by a new INS experiment, which reports a magnetic-field-induced phase transition in $\mathrm{YbMgGaO}_4$~\cite{William2019}. As the magnetic field increases, the peak of the magnetic susceptibility moves from the M point to the K point, as shown in Figs.~\ref{Fig:fig1} (a) and (b), implying that the ground state changes from a state with stripe-like correlations to a state with 120$^{\circ}$ antiferromagnetic (AFM)-like correlations (see Figs.~\ref{Fig:fig1} (c) and (d)). Ref.~\cite{William2019} showed that the C model cannot produce this change. Then a natural question arises. Can the A and B models produce this change? If not, can we find another homogeneous model to simultaneously describe this change at low fields and spin excitations at high fields?

To answer these questions, I carefully study the homogeneous spin Hamiltonian, which is extensively used to describe $\mathrm{YbMgGaO}_4$. 
Using the classical Monte Carlo method, the cluster mean-field theory, and exact diagonalization, the A, B, and C models are examined at different magnetic fields.
The results show that the stripe-like correlations do not vanish at low fields~\cite{Supplement}. Then I study the phase diagram using the cluster mean-field theory and exact diagonalization, and find that the magnetic field can change the ground state from a spin liquid state to a 120$^\circ$ AFM state, as observed in experiments. The cluster mean-field theory results and exact diagonalization results show that the next-nearest neighbor interaction should be small to produce this field-induced phase transition.
I analyze a parameter space relevant to $\mathrm{YbMgGaO}_4$ and calculate the spin excitation energy in that region using the linear spin-wave theory. Comparing to the INS data, I find that the homogeneous model in that parameter space can simultaneously describe the spin properties of $\mathrm{YbMgGaO}_4$ at both low and high fields.

\section{Model and method}
The homogeneous spin Hamiltonian~\cite{LiuPRB2016,dongPRB2016,Yuesheng2016PRL,YaodongPRB2017,Luo2017PRB,Parker2018PRB}, describing $\mathrm{YbMgGaO}_4$, is given by
\begin{eqnarray}\label{eq:hamiltonian}
H&=&\sum_{\langle ij\rangle} \Big\{ J_1^{zz} S_i^zS_j^z+J_1^{\pm}\left(S_i^+S_j^-+S_i^-S_j^+ \right) \nonumber\\
&&+J_1^{\pm\pm}\left(\gamma_{ij}S_i^+S_j^++\gamma_{ij}^{*}S_i^-S_j^-\right)\nonumber\\
&& -\frac{iJ_1^{z\pm}}{2}\left[\left(\gamma_{ij}^{*}S_i^+ - \gamma_{ij}S_i^- \right)S_j^z  + S_i^z\left(\gamma_{ij}^{*}S_j^+ - \gamma_{ij}S_j^- \right)  \right]\Big\}  \nonumber\\
&& + \sum_{\langle\langle ij\rangle\rangle} \left[ J_2^{zz} S_i^zS_j^z+J_2^{\pm}\left(S_i^+S_j^-+S_i^-S_j^+ \right) \right]\nonumber\\
&&-\sum_{i}\left[g_{\perp}\mu_0\mu_{B}(H_xS_i^x+H_yS_i^y)+g_{\parallel}\mu_0\mu_{B}H_zS_i^z \right],
\end{eqnarray}
where $\langle \cdots \rangle$ and $\langle\langle \cdots \rangle\rangle$ run over the nearest and next-nearest neighbors. $S_i^{\pm}=S_i^{x}\pm iS_i^{y}$ and $\gamma_{ij}=\gamma_{ji}=1,e^{i2\pi/3},e^{-i2\pi/3}$ are the phase factors for the bond $ij$ along the three principal directions~\cite{Parker2018PRB}. $J_1^{zz}$ and $J_1^{\pm}$ terms constitute the XXZ model; $J_1^{z\pm}$ and $J_1^{\pm\pm}$ are the spin-orbit interactions; $J_2^{zz}$ and $J_2^{\pm}$ are the next-nearest neighbor exchange interactions. To reduce the number of variables, I set $J_2^{zz}/J_1^{zz}=J_2^{\pm}/J_1^{\pm}=J_2/J_1$, following Ref.~\cite{ZhangPRX}. $H_x$, $H_y$, and $H_z$ are the magnetic field along the x, y, and z directions, respectively.

\begin{figure}[b]
\center\includegraphics[width=0.95\columnwidth]{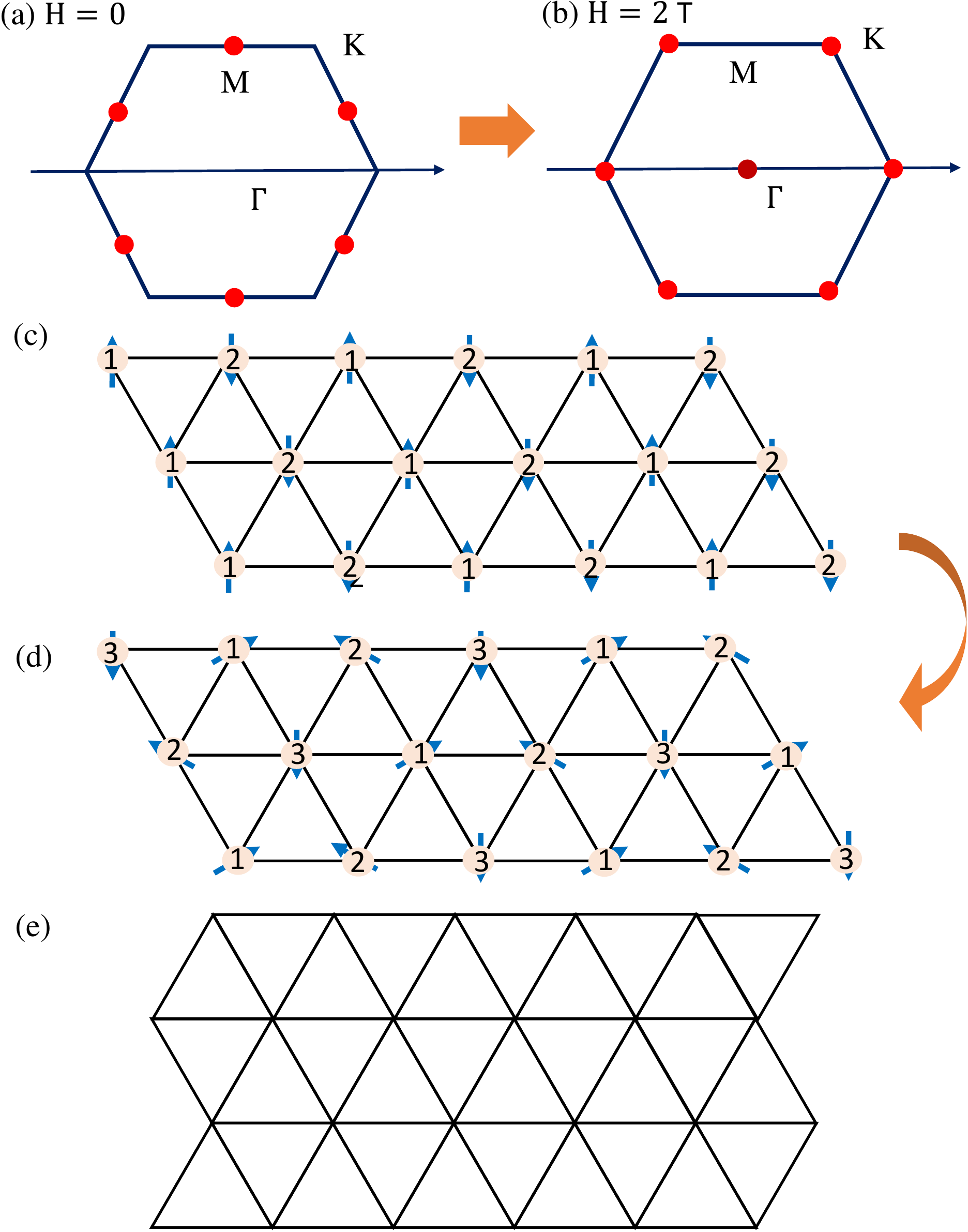}
\caption{\label{Fig:fig1} Schematic of the phase transition induced by the magnetic field in $\mathrm{YbMgGaO}_{4}$. In panels (a) and (b), red dots show the peak position of the static magnetic susceptibility in the Brillouin zone at the magnetic field $H_z=0$ and $H_z=2$ T. Panels (c) and (d) show the corresponding spin configurations in the real space. The number represents the index of atoms in one unit cell. Panel (e) shows a $6\times 4$ cluster for exact diagonalization simulations.}
\end{figure}

\begin{figure}[t]
\center\includegraphics[width=\columnwidth]{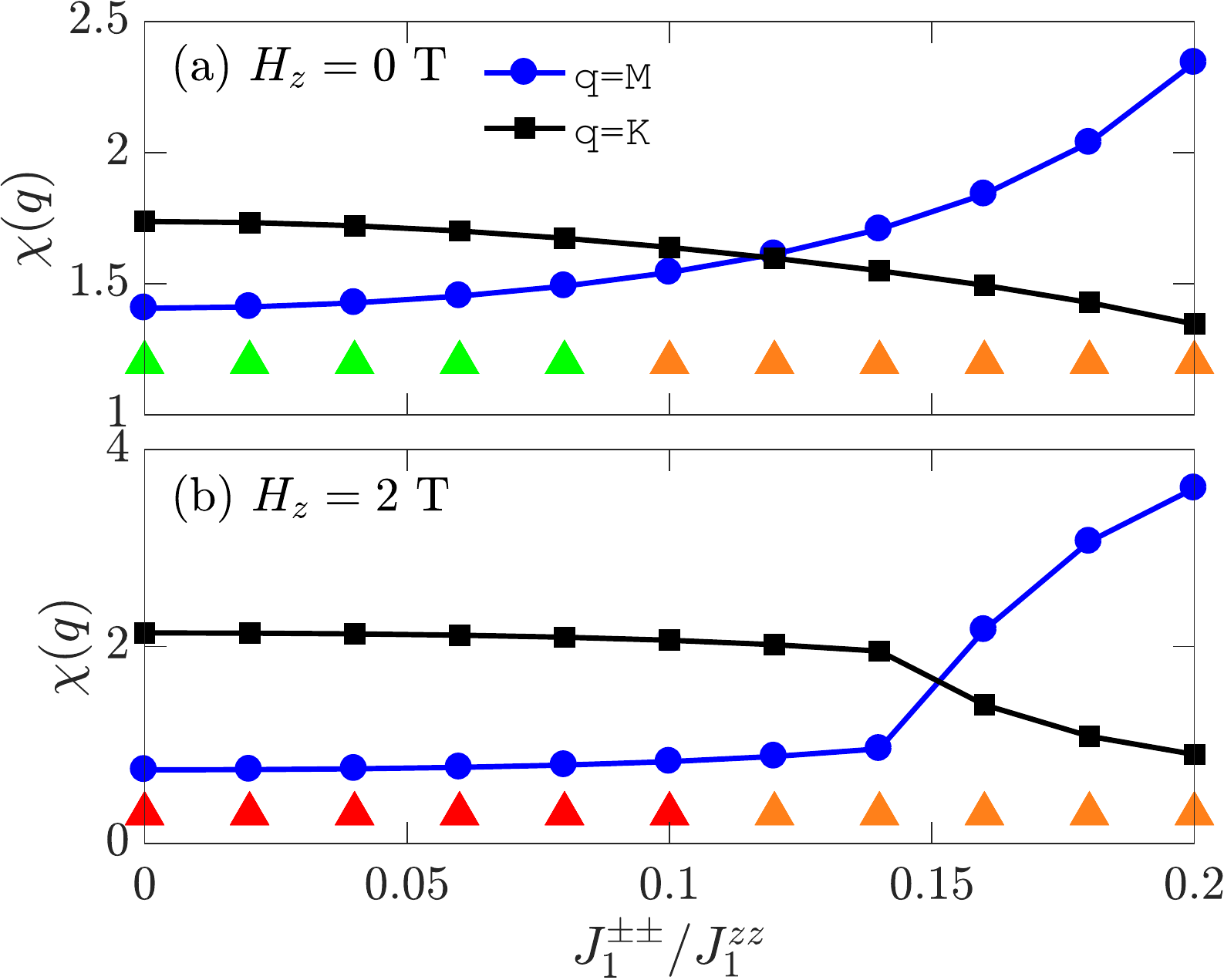}
\caption{\label{Fig:fig2} The spin correlation function $\chi(q)$ as a function of $J_1^{\pm\pm}$, evaluated by exact diagonalization. Panels (a) and (b) plot results for $H_z=0$ and $H_z=2$ T, respectively. Here, $M=(0,\frac{2\pi}{\sqrt{3}\alpha})$ and $K=(\frac{2\pi}{3\alpha},\frac{2\pi}{\sqrt{3}\alpha})$, where $\alpha$ is the lattice constant. Cluster mean-field theory calculations were performed at the triangular points. Green, red, and orange triangles represent the spin liquid state, the $120^{\circ}$ AFM state, and the stripe state, respectively. Here, $J_1^{zz}=0.162$ meV, $J_1^{\pm}/J_1^{zz}=0.66$, $J_1^{z\pm}=0$, $J_2/J_1=0.05$, and $g_{\parallel}=3.72$. }
\end{figure}

\begin{figure}[t]
\center\includegraphics[width=\columnwidth]{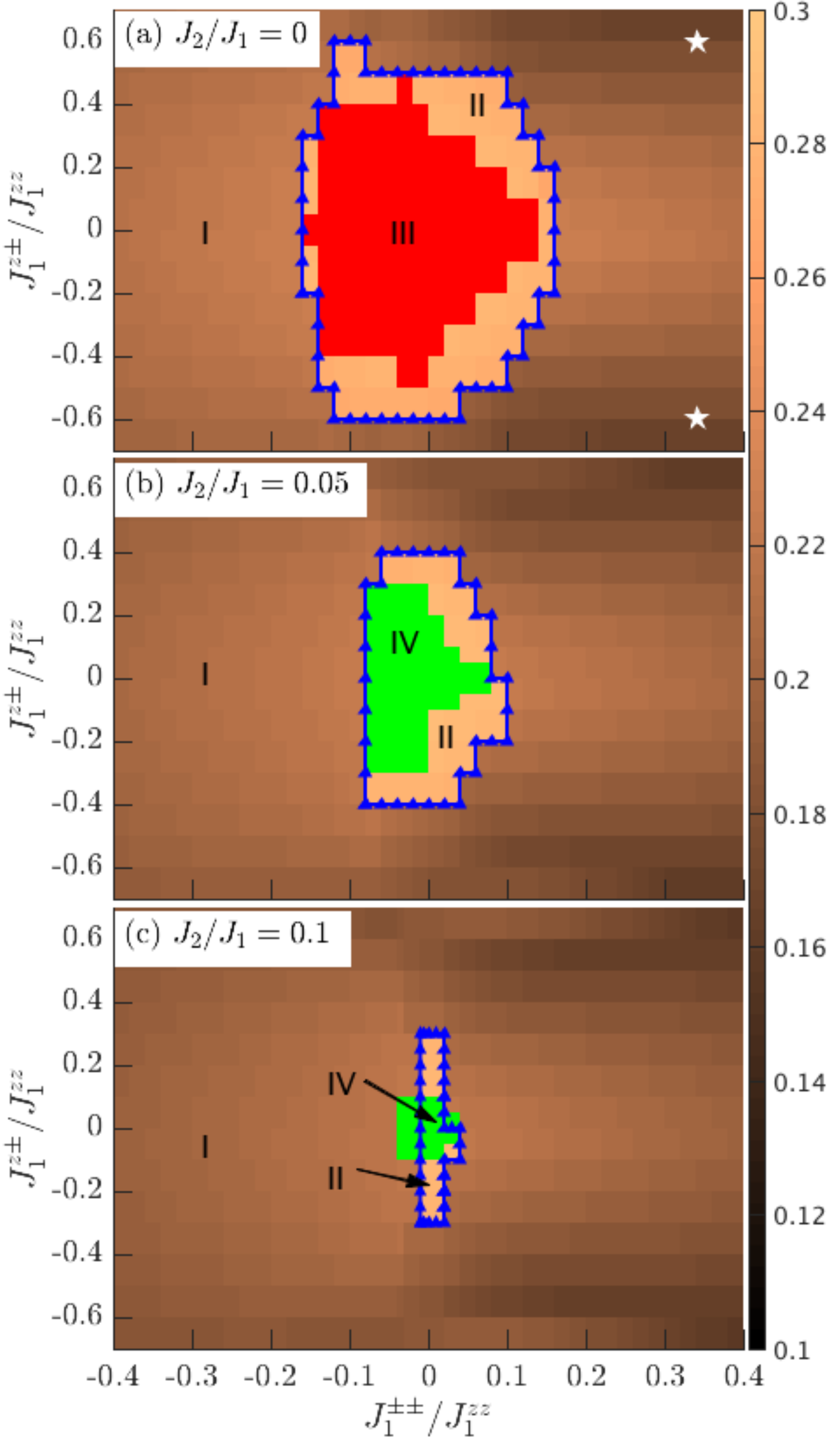}
\caption{\label{Fig:fig3} The phase diagram in the plane of $J_1^{\pm\pm}$ and $J_1^{z\pm}$. Panels (a), (b), and (c) show results for $J_2/J_1=0$, 0.05, and 0.1, respectively. The z-axis represents the uniform magnetization at $H_z=2$ T. The blue dotted-line represents the phase boundary between the stripe and the $120^{\circ}$ AFM states at $H_z=2$ T. The stripe state is inside the boundary and the $120^{\circ}$ AFM state is outside the boundary. The red and green regions correspond to the $120^{\circ}$ AFM and the spin liquid states at $H_z=0$, respectively. The white star shows the location of the B model in the phase diagram.
Here, $J_1^{zz}=0.164$ meV, $J_1^{\pm}=0.108$ meV and $g_{\parallel}=3.72$.}
\end{figure}

I study this Hamiltonian using the cluster mean-field theory and exact diagonalization.
The cluster mean-field theory is a self-consistent approach that exactly treats interactions inside the cluster and treats interactions between clusters at the mean-field level~\cite{YamaotoPRB2009,YamamotoPRL2014,YamaotoPRB2017,Ren_2014}. In the cluster mean-field theory, the lattice Hamiltonian is rewritten as $H=\sum_{C_\gamma} H_{C_\gamma} + \sum_{C_\gamma,C_\lambda} H_{C_\gamma  C_\lambda}$, where $H_{C_\gamma}$ describes interactions inside the cluster $C_\gamma$, and $H_{C_\gamma  C_\lambda}$ describes interactions between two neighboring clusters. The total energy of the cluster is given by $E_c=E-E_b/2$,
where $E$ is the energy of the cluster $C$, and $E_b$ is the energy of interactions between two clusters, which is given by $\sum_{\langle ij\rangle}J_{ij}\langle S_i \rangle\langle S_j\rangle$.

Throughout this work, I focus on the $120^{\circ}$ AFM state and the stripe state. The $120^{\circ}$ AFM state has three sites in one unit cell (see Fig.~\ref{Fig:fig1} (d)), and the stripe state has two sites in one unit cell (see Fig.~\ref{Fig:fig1} (c)). For each parameter set, I performed simulations for both states at zero temperature. The ground state is determined by the one with a smaller $E_c$.
Besides these two states, a spin liquid state is observed, in which the uniform magnetization $\langle S^{\sigma}\rangle \approx 0$, where $\sigma=x$, $y$, and $z$~\cite{Supplement}.

In this work, cluster mean-field theory is performed on a $6\times3$ cluster as shown in Figs.~\ref{Fig:fig1} (c) and (d). I examine the cluster mean-field theory results at $J_1^{\pm}/J_1=0.66$, $J_1^{z\pm}$=0 and $J_1^{\pm\pm}=0$. It shows that the spin liquid state is located in the region $J_2/J_1\in [0.04,0.2]$, consistent with the density matrix renormalization group theory results ($J_2/J_1\in [0.06,0.17]$)~\cite{ZhuPRB2015,Supplement}. Besides, I compare the cluster mean-field theory results to exact diagonalization results in Fig.~\ref{Fig:fig2}. Here, $J_1^{zz}=0.162$ meV, $J_1^{\pm}/J_1^{zz}=0.66$, $J_1^{z\pm}=0$, $J_2/J_1=0.05$, and $g_{\parallel}=3.72$.
Figure~\ref{Fig:fig2} plots the spin correlation function $\chi(q)$ obtained from exact diagonalization, which is performed on a $6\times 4$ lattice with the periodic boundary condition, as shown in Fig.~\ref{Fig:fig1} (e). 
The spin correlation function $\chi(q)$ is given by $\chi(q)=\sum_{\sigma} \langle S^{\sigma}_q S^{\sigma}_{-q} \rangle$, where $S_q^{\sigma}$ is the Fourier transformation of $S_i^{\sigma}$.
On the $6\times 4$ lattice, $\chi(q=M)$ has the maximum value for the stripe state, and $\chi(q=K)$ has the maximum value for the $120^{\circ}$ AFM state, where $M=(0,\frac{2\pi}{\sqrt{3}\alpha})$, $K=(\frac{2\pi}{3\alpha},\frac{2\pi}{\sqrt{3}\alpha})$, and $\alpha$ is the lattice constant. 
Cluster mean-field theory calculations were performed at the triangular dots in Fig.~\ref{Fig:fig2}.
Green, red and orange triangles represent the spin liquid state, the $120^{\circ}$ AFM state, and the stripe state, respectively.
Exact diagonalization results show that the ground state is the $120^{\circ}$ AFM state when $J_1^{\pm\pm}/J_1^{zz}<0.12$ and $H_z=0$, while $|\chi(K)-\chi(M)|$ is small. As $J_1^{\pm\pm}$ increases, the ground state becomes the stripe state, and $\chi(M)$ increases rapidly.
Due to the finite size effect, the spin liquid state, predicted by cluster mean-field theory, is absent here. In spite of this difference, both results show that the dominant spin correlation becomes the stripe-like correlation near $J_1^{\pm\pm}/J_1^{zz}=0.11$.
At $H_z=2$ T, exact diagonalization results show that the critical value of the phase transition is about 0.15, which is 0.03 larger than the value predicted by cluster mean-field theory. All these results imply that the critical value of changing spin correlations predicted by cluster mean-field theory is close to the value predicted by other techniques.

\section{Results} 
\subsection{Results of the cluster mean-field theory}
In this work, the phase diagrams of the Hamiltonian at $H_z=0$ and $H_z=2$ T are studied. At $H_z=0$, there are three states in the phase diagram: the $120^{\circ}$ AFM state, the spin liquid state, and the stripe state. 
The red patch in the phase diagram represents the $120^{\circ}$ AFM state at $H_z=0$, and the green patch represents the spin liquid state. The stripe state is not labeled. 
At $H_z=2$ T, the spin liquid state disappears, and there are only two states in the phase diagram: the stripe state and the $120^{\circ}$ AFM state. These two states are separated by the blue dotted-line in the phase diagram. The $120^{\circ}$ AFM state is inside the blue dotted-line, and the stripe state is outside the blue dotted-line.

\begin{figure}[t]
\center\includegraphics[width=\columnwidth]{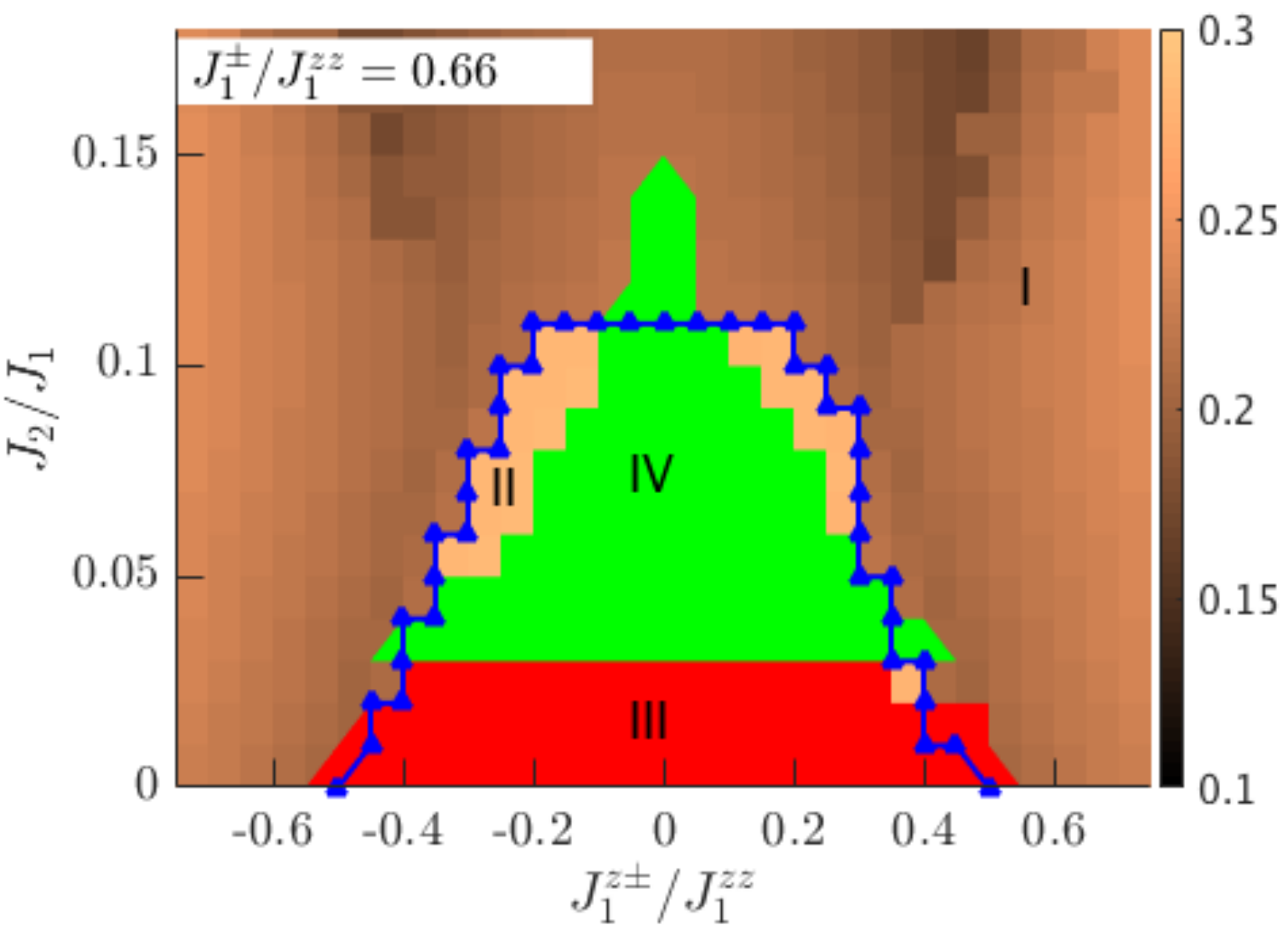}
\caption{\label{Fig:fig4} The phase diagram in the plane of $J_1^{z\pm}$ and $J_2$ at $J_1^{\pm}/J_1^{zz}=0.66$. The blue dotted-line is the phase boundary between the stripe and the $120^{\circ}$ AFM states at $H_z=2$ T. The red and green regions correspond to the $120^{\circ}$ AFM and spin liquid states at $H_z=0$, respectively.}
\end{figure}

Figure~\ref{Fig:fig3} plots the phase diagrams as a function of $J_1^{\pm\pm}$ and $J_1^{z\pm}$ for $J_2/J_1=0$, 0.05, and 0.1, respectively. Here, $J_1^{zz}=0.164$ meV, $J_1^{\pm}=0.108$ meV, and $g_{\parallel}=3.72$. The z-axis represents the uniform magnetization $\langle \hat{S}^z\rangle$ at $H_z=2$ T. 
At $J_2/J_1=0$ and $H_z=0$ (see Fig.~\ref{Fig:fig3} (a)), the $120^{\circ}$ AFM state (red patch labeled as ``III") is located at the center of the phase diagram.
The phase boundary of the $120^{\circ}$ AFM state is asymmetric about $J_1^{\pm\pm}=0$ as predicted in previous classical Monte Carlo and linear spin-wave studies~\cite{dongPRB2016}.
At $H_z=2$ T, the phase boundary (blue dotted-line) for the $120^{\circ}$ AFM state is slightly broadened, leading to the formation of region II.
In region II, a phase transition from the stripe state to the $120^{\circ}$ AFM state occurs as the magnetic field increases. I label the B model as a white star in the phase diagram, outside of the blue dotted-line. The field-induced phase transition is absent for the B model is because $J_1^{\pm\pm}$ is too large.

\begin{figure}[t]
\center\includegraphics[width=\columnwidth]{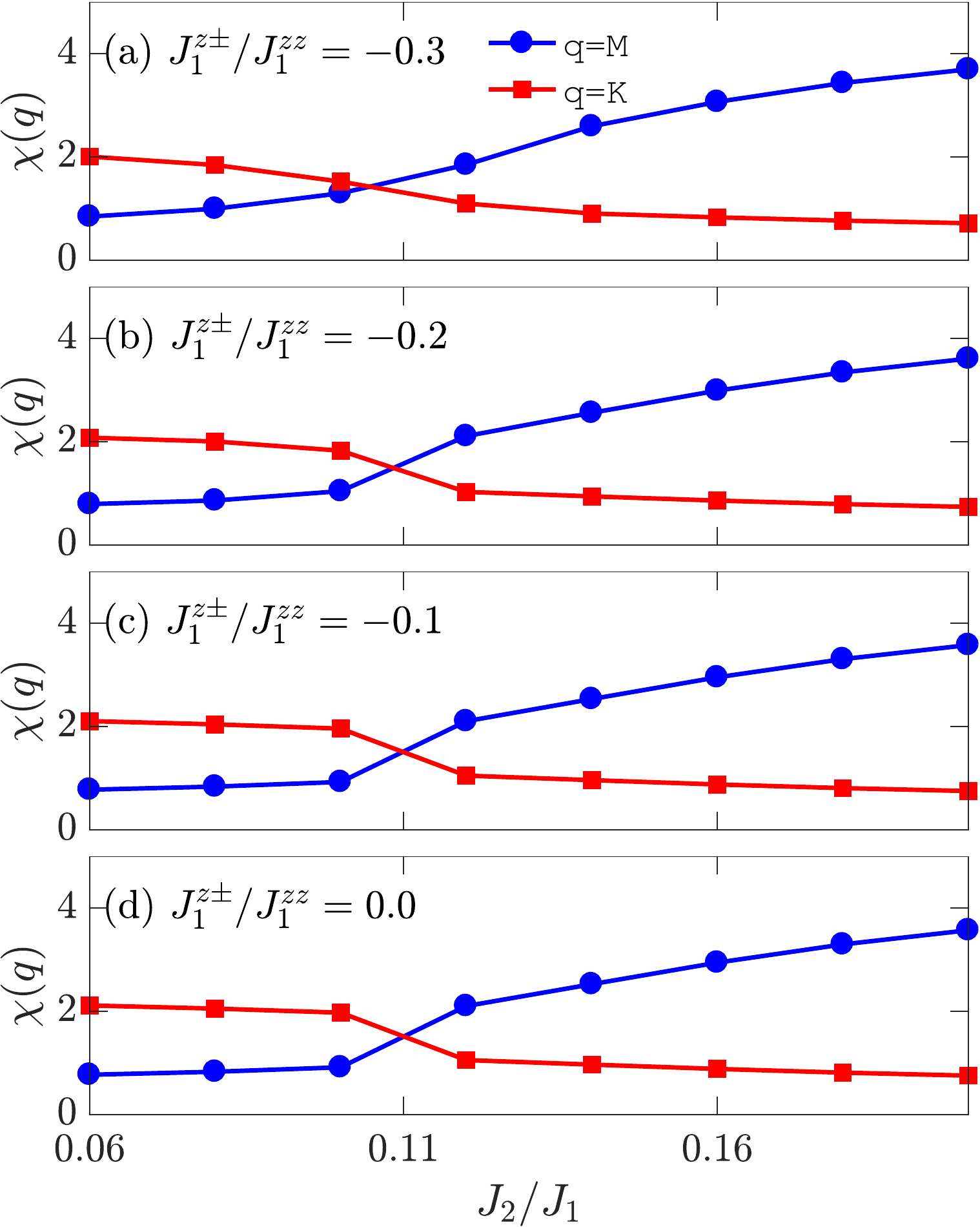}
\caption{\label{Fig:fig5}  The spin correlation function $\chi(q)$ as a function of $J_2/J_1$, evaluated by exact diagonalization. Here, $M=(0,\frac{2\pi}{\sqrt{3}\alpha})$ and $K=(\frac{2\pi}{3\alpha},\frac{2\pi}{\sqrt{3}\alpha})$, where $\alpha$ is the lattice constant. The other parameters are $J_1^{\pm}/J_1^{zz}=0.66$, $J_1^{\pm\pm}$=0, $H_z=2$ T, and $g_{\parallel}=3.72$.
}
\end{figure}

\begin{figure}[t]
\center\includegraphics[width=\columnwidth]{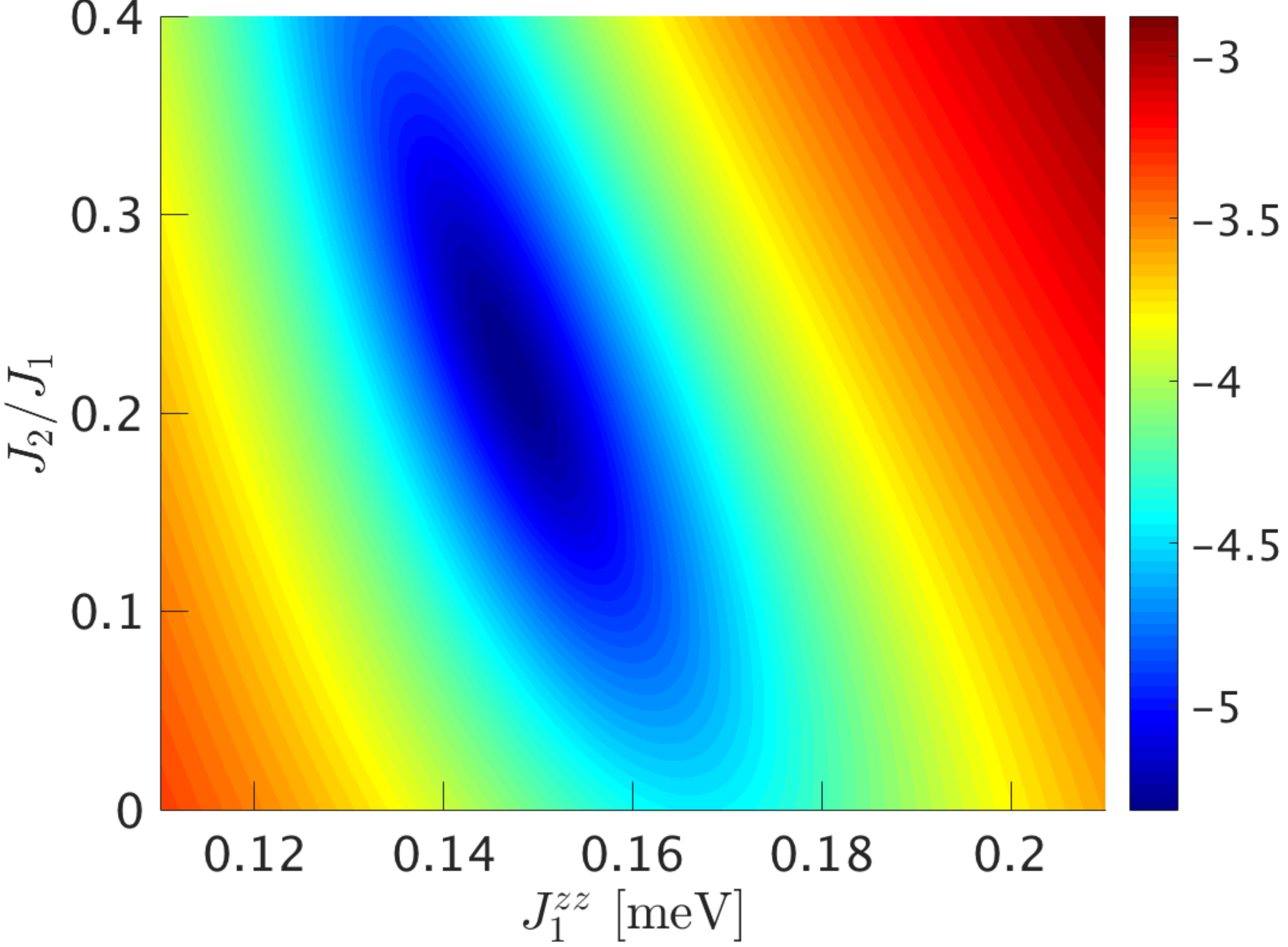}
\caption{\label{Fig:fig6}  Deviation $\mathrm{log}(R)$ of the experimental spin-wave energies~\cite{ZhangPRX} from theoretical values as a function of interactions $J_1^{zz}$ and $J_2$.}
\end{figure}

At $J_2/J_1=0.05$ and $H_z=0$ (see Fig.~\ref{Fig:fig3} (b)), the $120^{\circ}$ AFM state is replaced by the spin liquid state (green patch) at the center of the phase diagram. This result is consistent with previous DMRG predictions that a small $J_2$ induces a spin liquid state~\cite{ZhuPRB2015}. At $H_z=2$ T, the area of the $120^{\circ}$ AFM state decreases compared to that at $J_2/J_1=0$. Continuing to increase $J_2/J_1$ to 0.1, the region of the $120^{\circ}$ AFM state becomes very small.
The critical value of $J_1^{\pm\pm}$ decreases faster than the critical value of $J_1^{z\pm}$ as $J_2$ increases.
These results imply that both the next-nearest neighbor interaction and the SOC $J_1^{\pm\pm}$ suppress the field-induced phase transition.

\begin{figure*}[t]
\center\includegraphics[width=\textwidth]{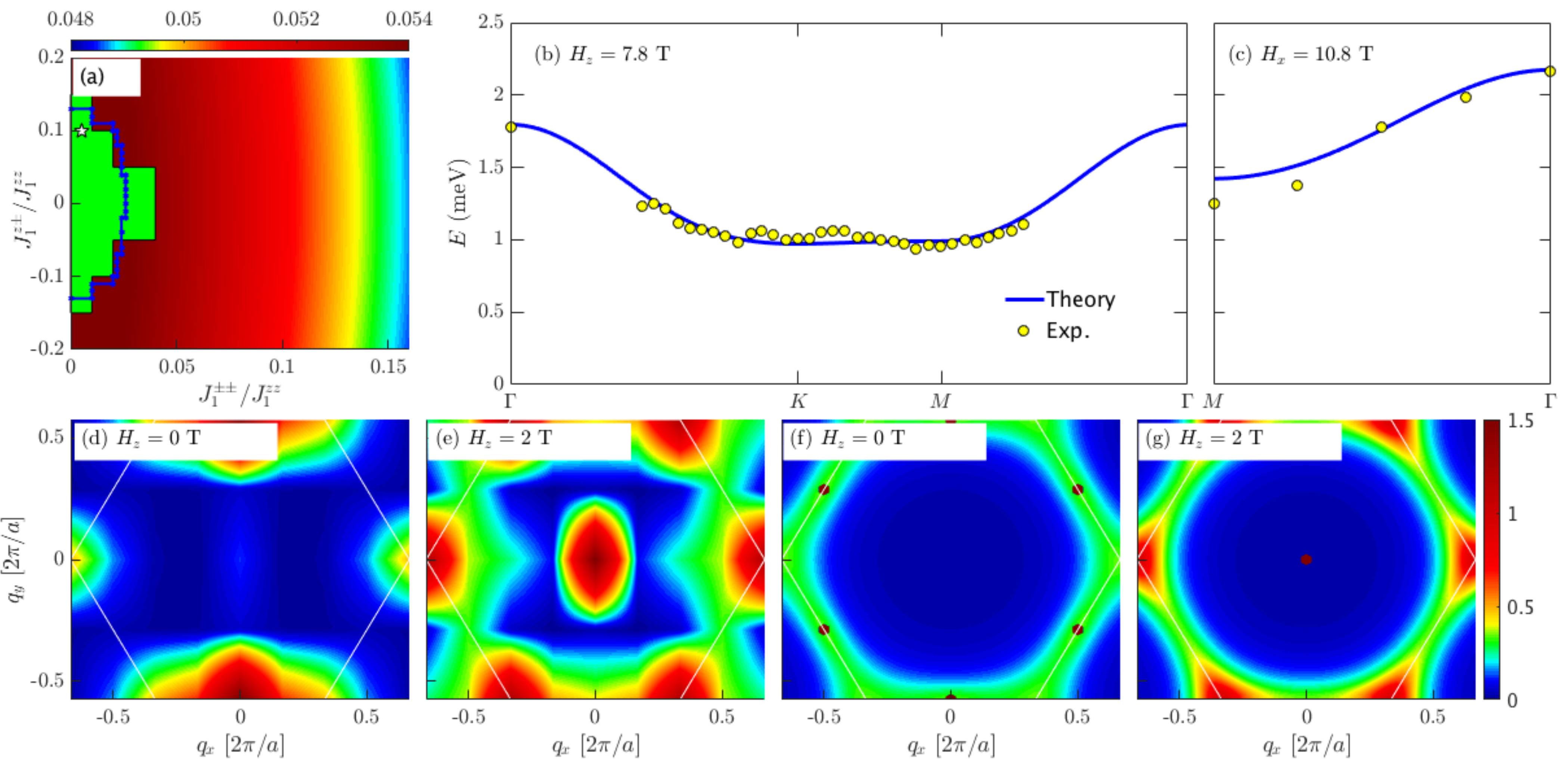}
\caption{\label{Fig:fig7} Panel (a) shows deviation $R$ of the experimental spin-wave energies from the theoretical values as a function of interactions $J_1^{z\pm}$ and $J_1^{\pm\pm}$. Here, $J_1^{zz}=0.158$ meV, $J_1^{\pm}=0.09155$ meV, and $J_2/J_1=0.1$. The green region near $J_1^{\pm\pm}=0$ corresponds to the spin liquid state at $H_z=0$, and the blue dotted-line represents the phase boundary at $H=2$ T. Panels (b) and (c) plot spin-wave energies in the momentum space. Experimental results (yellow dots) are obtained from Ref.~\cite{ZhangPRX}. Panels (d) and (e) plot the spin correlation function $\chi(q)$, which is obtained from the exact diagonalization. Panels (f) and (g) plot the spin correlation function $\chi(q)$ obtained from the classical Monte Carlo simulations.}
\end{figure*}

To further explain the effect of $J_2$, I examine the phase diagram in the plane of $J_2$ and $J_1^{z\pm}$ in Fig.~\ref{Fig:fig4}. Here, $J_1^{\pm\pm}=0$ and the other parameters are the same as those in Fig.~\ref{Fig:fig3}.
At $H_z=0$, there is a 120$^\circ$ AFM state (red patch) when $J_2/J_1<0.03$. This AFM state becomes the spin liquid state (green patch) as $J_2$ increases. At $H_z=2$ T, the phase boundary (blue dotted-line) goes through the spin liquid region (green patch) at $J_2/J_1=0.11$, implying that a field-induced phase transition between the spin liquid state and the 120$^\circ$ AFM state occurs as the magnetic field is applied. There is a maximal value of $J_2/J_1$ ($\approx 0.11$), above which this field-induced phase transition cannot occur.
The ratio of $J_1^{\pm}/J_1^{zz}$ has a tiny effect on this maximal value (see Ref.~\cite{Supplement}).

\subsection{Results of the exact diagonalization}
To corroborate the result that $J_2$ should be small to make the field-induced phase transition occur, I perform exact diagonalization calculations on the $6\times4$ lattice. 
To compare to Fig.~\ref{Fig:fig4}, parameters are set as $J_1^{zz}=0.164$ meV, $J_1^{\pm}=0.66$, $H_z$=2 T, and $g_{\parallel}=3.72$. Figure~\ref{Fig:fig5} plots the spin correlation function $\chi(q)$ for different $J_1^{z\pm}/J_1^{zz}$. Figure~\ref{Fig:fig5} (a) shows that $\chi(K)$ decreases and $\chi(M)$ increases as $J_2/J_1$ increases. $\chi(K)$ and $\chi(M)$ cross at $J_2/J_1=0.1$. As $J_1^{z\pm}$ increases, the intersection moves toward $J_2/J_1=0.11$. All these results imply that the dominant spin correlations become stripe-like when $J_2/J_1>0.11$.

The results from exact diagonalization are consistent with the results from cluster mean-field theory. Both results show that the dominant spin correlations change near $J_2/J_1=0.1$ at $H_z=2$ T. 
Keeping these results in mind, I next analyze experimental results and search for a good model for $\mathrm{YbMgGaO}_4$. Note that the phase diagram could be modified on an infinite lattice and a good model for $\mathrm{YbMgGaO}_4$ could be different.

\subsection{Relevance to $\mathrm{YbMgGaO}_4$}
I study the Hamiltonian using the linear spin-wave theory.
At high fields, the spin excitation energy at the $\Gamma$ point is given by
\begin{eqnarray}
E(H_z)=g_{\parallel}\mu_0\mu_{B}H_z-2x
\end{eqnarray}
and
\begin{eqnarray}
E(H_x)=\sqrt{\left(g_{\perp}\mu_0\mu_{B}H_x+x\right)^2-x^2},
\end{eqnarray}
where $x=\frac{3}{2}(J_1^{zz}+J_2^{zz})-3(J_1^{\pm}+J_2^{\pm})$. By fitting these two equations to the time-domain terahertz spectra obtained from Ref.~\cite{ZhangPRX}, I obtain  $g_{\parallel}=3.8$, $g_{\perp}=3.55$, and $x=-0.0414$~\cite{Supplement}. This fitting strategy is different from that in Ref.~\cite{ZhangPRX}.
Considering results shown in Fig.~\ref{Fig:fig3} and Fig.~\ref{Fig:fig4}, I calculate the momentum-dependent spin excitation energy $E_\mathrm{sw}({\bf q})$ in the parameter space \{$J_1^{zz}\in[0.11,0.21]$, $J_2/J_1\in[0,0.4]$, $J_1^{\pm\pm}/J_1^{zz}\in[-0.2,0.2]$, and $J_1^{z\pm}/J_1^{zz}\in[-0.6, 0.6]$\}. The deviation $R$ of $E_\mathrm{sw}({\bf q})$ from the INS data $E_\mathrm{exp}({\bf q})$~\cite{ZhangPRX} is defined as $R=\frac{1}{\sqrt{L}}\sum_{i}\sqrt{\left(E_\mathrm{sw}({\bf q}_i) - E_\mathrm{exp}({\bf q}_i)\right)^2}$, where $L$ is the total number of momenta obtained from experiments. Figure~\ref{Fig:fig6} plots $\mathrm{log}(R)$ as a function of $J_1^{zz}$ and $J_2/J_1$. Here, the values of $J_1^{\pm\pm}$ and $J_1^{z\pm}$ are determined by producing the smallest $R$ at a given $J_1^{zz}$ and $J_2/J_1$. The minimal $R$ is located at $J_1^{zz}=0.146$ meV and $J_2/J_1=0.22$. There is a small difference between the C model and my results because only a part of the experimental data shown in Ref.~\cite{ZhangPRX} is used. That part is plotted in Figs.~\ref{Fig:fig7} (b) and (c). Despite a small difference, both the C model and my results show that $J_2/J_1\approx 0.2$ for the best fitting.

The INS data of $\mathrm{YbMgGaO}_4$ has a broad spectrum, and the energy of the peak is not well defined~\cite{Paddison2017}. Therefore, the best fitting results may not provide the best model for $\mathrm{YbMgGaO}_4$. A good model for $\mathrm{YbMgGaO}_4$ could be one that produces a field-induced transition while giving a small deviation $R$.
Consequently, I choose a large allowed value of $J_2$ ($J_2/J_1=0.1$) and carefully analyze results. Figure~\ref{Fig:fig7} (a) plots $R$ as a function of interactions $J_1^{z\pm}$ and $J_1^{\pm\pm}$. Here, $J_1^{zz}=0.158$ meV because it produces the smallest $R$ at a given $J_1^{z\pm}$, $J_1^{\pm\pm}$, and $J_2/J_1$. 
The green region near $J_1^{\pm\pm}=0$ corresponds to the spin liquid state at $H_z=0$, and the blue dotted-line is the phase boundary at $H_z=2$ T.
$R$ decreases slowly as $J_1^{\pm\pm}$ increases. I choose one point (white star) inside the blue dotted-line in Fig.~\ref{Fig:fig7} (a) and plot the momentum-dependent spin excitation energies in Figs.~\ref{Fig:fig7} (b) and (c). The parameters for the white star are 
$J_1^{zz}=0.158$ meV, $J_1^{\pm}=0.092$ meV, $J_1^{\pm\pm}/J_1^{zz}=0.005$, $J_1^{z\pm}/J_1^{zz}=0.1$, and $J_2/J_1=0.1$. 
Yellow dots in Figs.~\ref{Fig:fig7} (b) and (c) are experimental results obtained from Ref.~\cite{ZhangPRX}.
The theoretical line and the experimental data almost overlap with a small deviation $R\approx 0.054$. Note that all parameters inside the blue dotted-line in Fig.~\ref{Fig:fig7} (a) are good enough to describe $\mathrm{YbMgGaO}_4$.

Next, using the parameters of the white star, I study the spin correlation function $\chi(q)$ at $H_z=0$ and $H_z=2$ T.
Figures~\ref{Fig:fig7} (d) and (e) plot the exact diagonalization results. Due to the small size of the lattice, a linear interpolation is used here. At $H_z=0$, the ground state is the stripe state, and the intensity of $\chi(M)$ is the maximum. At $H_z=2$ T, the ground state has a $120^{\circ}$ AFM correlation, and the intensity of $\chi(K)$ is strong. Besides, I also perform classical Monte Carlo simulations for the same parameters on a $48\times 48$ lattice. Results are shown in Figs.~\ref{Fig:fig7} (f) and (g). A similar field-induced phase transition is observed.

\section{Summary}
I have carefully studied the spin Hamiltonian to understand the spin properties in $\mathrm{YbMgGaO}_{4}$. I calculate the phase diagram of the Hamiltonian using the cluster mean-field theory and exact diagonalization. The results show that the magnetic field can induce phase transitions between the stripe state ( or the spin liquid state) and the 120$^{\circ}$ AFM state.
These phase transitions are suppressed by the next-nearest neighbor interaction and the SOC $J_1^{\pm\pm}$. There is a maximal value of $J_2/J_1$, above which the field-induced phase transition cannot occur. By analyzing experimental results, I find a parameter space, where the field-induced phase transition can occur and the INS data can be reproduced with a deviation of $0.054$.

The good model for $\mathrm{YbMgGaO}_{4}$ found in this work is consistent with the best model shown in Ref.~\cite{William2019}, which is obtained from the classical Monte Carlo method. Both results show that the next-nearest neighbor interaction in $\mathrm{YbMgGaO}_{4}$ should be small ($J_2/J_1\approx 0.1$). 
The study with a quantum technique in this work and the study with a classical technique in Ref.~\cite{William2019} complement each other. Moreover, the phase diagrams shown in this work are helpful to understand other spin liquid candidates, including $\mathrm{NaYbO}_2$~\cite{RanjithPRB,DingPRB2019},  $\mathrm{NaYbS}_2$~\cite{SarkarPRB}, and $\mathrm{NaYbSe}_2$~\cite{RanjithPRB2019}.  

In this work, the phase diagram is studied on a $6\times 3$ cluster. It is unclear how the phase diagram is modified on the infinite lattice. Also, it maybe interesting to study how disorder changes this phase diagram. To fully understand $\mathrm{YbMgGaO}_{4}$, a further study is required.

\begin{acknowledgments}
I thank C. D. Batista and A. M. Samarakoon for early
discussions. This work was supported by the National Science
Foundation (NSF) under Grant No. DMR-1606348 and by the Simons Foundation via the Simons Collaboration on the Many-Electron Problem. 
This work was also supported by the U.S. Department of Energy, Office of Basic Energy Sciences, Materials Sciences and Engineering Division.
This work used resources of the Extreme Science and Engineering Discovering Enviroment (XSEDE) under Grant
No. TG-DMR130036 and No. TG-DMR190090. A part of CPU time
was provided by the University of
Tennessee and Oak Ridge National Laboratory Joint Institute
for Computational Sciences (http://www.jics.utk.edu).
\end{acknowledgments}

\bibliography{main}

\end{document}